\documentclass{jpp}

\usepackage{graphicx}
\usepackage[colorlinks=true, citecolor=blue, linkcolor=blue]{hyperref} 

\usepackage{etoolbox}

\usepackage{amsmath}
\usepackage{bm}
\usepackage{relsize}

\usepackage{caption}
\usepackage{subcaption}
\usepackage{xcolor}
\usepackage{multirow}

\usepackage[utf8]{inputenc}
\usepackage[T1]{fontenc}
\usepackage{algorithm}
\usepackage{algpseudocode}

\newcommand{\dx}{\Delta x}
\newcommand{\dc}{\Delta c}

\title{Applications of Poincaré Boundary Condition for 3D Ideal MHD Equilibrium and Optimization}

\shorttitle{Applications of Poincaré Boundary Condition for 3D Ideal MHD}
\shortauthor{Y. G. Elmacioglu et al.} 

\author{Yigit Gunsur Elmacioglu\aff{1}
  \corresp{\email{yigit.elma@princeton.edu}}, Dario Panici\aff{1}, Rory Conlin\aff{2}, Daniel Dudt\aff{3},
Egemen Kolemen\aff{1}\corresp{\email{ekolemen@princeton.edu}}}

\affiliation{
    \aff{1}Princeton University, Princeton, New Jersey 08544, USA
    \aff{2}Max Planck Institute for Plasma Physics, Greifswald, 17491, DE
    \aff{3}Thea Energy, USA
}

\begin{document}
\maketitle

\begin{abstract}
Stellarator ideal magnetohydrodynamic (MHD) codes that assume nested flux surfaces such as \texttt{VMEC} and \texttt{DESC} solve the equilibrium problem by prescribing the total toroidal magnetic flux, plasma profiles and the last closed flux surface (LCFS), which analytically determines the unique field in vacuum, whereas at finite $\beta$ bifurcations and distinct equilibria sharing the same boundary have been reported in the literature. In this paper, we propose prescribing the Poincaré cross-section of the field at a single toroidal plane, and implement it in \texttt{DESC}, where the new condition enters only through the linear constraints and therefore costs no more than a fixed-LCFS solve. Fixing the geometry on one plane rather than on a full toroidal surface generally leaves more of the spectral coefficients free, and the ones it frees are those carrying the toroidal variation of the boundary flux surface, which a prescribed LCFS holds fixed at every toroidal angle; together these allow better-converged numerical solutions. We solve equilibria with the cross-section held fixed, starting either from the axisymmetric shape obtained by revolving that cross-section toroidally, or from an existing fixed-LCFS solution. In the latter case, the volume-averaged normalized force error falls by an order of magnitude while the configuration stays close to the original one, and re-solving the resulting boundary with the conventional fixed-LCFS solver recovers the same equilibrium. We further show that the Poincaré coefficients can be used directly as design variables by optimizing a quasi-helical configuration that maintains high-fidelity force balance throughout the process.


\end{abstract}

\section{Introduction}

Stellarators are a class of magnetic confinement fusion devices that use complex, three-dimensional magnetic fields to confine plasma. Unlike tokamaks, they do not require a large induced plasma current for confinement, which grants them inherent stability against major current-driven disruptions \citep{helanderTheoryPlasmaConfinement2014}. The intricate 3D nature of the stellarator magnetic field makes analytical solutions to the governing ideal magnetohydrodynamic (MHD) equations intractable \citep{freidbergIdealMHD2014}, except in simplified regimes like near-axis expansions or high aspect ratio. Consequently, progress in the stellarator field has been driven primarily by numerical simulations.

The vast design space of stellarators presents promising candidates for a viable fusion reactor, making optimization methods essential. Nested-flux-surface equilibrium codes like \texttt{VMEC} \citep{hirshmanSteepestdescentMomentMethod1983} and \texttt{DESC} \citep{dudtDESCStellaratorEquilibrium2020a, dudtDESCStellaratorCode2023, paniciDESCStellaratorCode2023, conlinDESCStellaratorCode2023a} are frequently used in these optimization studies because the assumption of nested magnetic flux surfaces makes the computation more efficient. While higher-fidelity codes that do not assume nested surfaces exist (e.g., \texttt{SPEC} \citep{hudsonComputationMultiregionRelaxed2012}, \texttt{HINT2} \citep{suzuki_development_2006}) and have been incorporated in the optimization studies \citep{landremanStellaratorOptimizationGood2021, baillodStellaratorOptimizationNested2022}, their computational cost currently limits their use in large-scale optimization. 

Equilibrium solvers typically leverage the toroidal periodicity of the system to represent flux surfaces spectrally with a double Fourier series. The equilibrium solution is then found by optimizing the spectral coefficients to minimize either the total MHD energy or the ideal MHD equilibrium force balance error, $||\mathbf{J}\times\mathbf{B}-\nabla p||$. This calculation requires a set of inputs: a pressure profile, a current or rotational transform profile, the total enclosed magnetic flux, and the shape of the last closed flux surface (LCFS)\footnote{We note that the computational boundary we supply might not correspond to the actual last closed flux surface of the system. However, we will keep referring to the boundary as LCFS for the lack of a better word.}, the toroidal boundary to which the magnetic field is tangent \citep{kruskalEquilibriumMagneticallyConfined1958}. The assumption underlying this construction, that the volume is covered everywhere by nested flux surfaces, is a strong one, and it is not satisfied by a general three-dimensional field: islands and chaotic regions are generic features, and their absence for a given LCFS is not guaranteed \citep{grad_toroidal_1967, garabedian_computational_2003}. Solvers of this class therefore do not guarantee an island-free field. They return the best nested-surface representation available at the given resolution, and the residual force error that they report is in part a measure of the field structure that such a representation cannot express \citep{lazerson_verification_2016}.

Once the solution is understood as an approximation to a field that is not globally nested, prescribing the LCFS becomes one of several ways to close the problem rather than a privileged one. Its strongest justification is in vacuum, where the physical field satisfies Laplace's equation and the boundary shape, together with the total enclosed magnetic flux, determines the solution uniquely. At finite $\beta$, which is the regime that stellarator optimization ultimately targets, no such guarantee exists. Distinct equilibria can share the same LCFS, as in the bifurcated tokamak states with an axisymmetric boundary and a three-dimensional helical core \citep{cooperTokamakMagnetohydrodynamicEquilibrium2010, paniciDeflationTechniquesStellarator2026}, where the magnetic axis acquires a helical excursion that the boundary shape does not reveal; other bifurcated equilibria have also been discussed in the literature \citep{garabedian_computational_2003, ilgisonisBifurcationEquilibriumCurrentcarrying2004, garabedianBifurcatedEquilibriaMagnetic2006}. The analytical justification of the LCFS boundary condition is therefore confined to the vacuum limit that optimization studies are moving away from.

In this paper we propose a novel alternative: prescribing the Poincaré cross-section of the field at a single toroidal plane, which we refer to as the Poincaré boundary condition. Prescribing the geometry on a single plane rather than on a full toroidal surface leaves more degrees of freedom, giving the equilibrium solver a larger space to search. Whenever the representation of the equilibrium, the 3D spectral basis of \texttt{DESC}, for instance, allows the cross-section to be written as a linear relation between the coefficients, the condition enters the solve only through the linear constraints and therefore the per-step cost is no more than that of a fixed-LCFS solve. This is an algorithmic choice rather than an analytical one, and we return in the next section to what it does and does not imply.

This paper is organized as follows. First, we will detail the numerical implementation of this new boundary condition within the \texttt{DESC} code. Next, we will demonstrate its versatility by exploring several use cases: solving for MHD equilibria, refining existing equilibria to reduce force errors, and finally, incorporating the Poincaré section shape as a variable in optimization studies.

\section{Numerical Model}
\texttt{DESC} is a pseudo-spectral code and represents the shape of the nested flux surfaces, $R$ and $Z$, as well as the poloidal stream function $\lambda$ by a 3D Fourier-Zernike basis. The computational coordinates $(\rho, \theta, \zeta)$, where $\rho=\sqrt{\psi/\psi_b}$ is the square root of the normalized toroidal flux, $\theta$ is an arbitrary poloidal angle and $\zeta$ is the toroidal angle, are mapped to the cylindrical coordinates $(R, \phi, Z)$ and to the straight field-line angle $\vartheta = \theta + \lambda$, in which the magnetic field can be written in terms of the surface geometry and the rotational transform alone. In this work, we take the toroidal angle $\zeta$ to be the same as the cylindrical angle $\phi$. A general function $f$ (which is either $R, Z$ or $\lambda$) is written as,

\begin{equation}
    f(\rho,\theta,\zeta) = \sum_{l,m}^{}\sum_{n}^{} f_{lmn} \mathcal{Z}_l^{m} (\rho, \theta) \mathcal{F}^n(\zeta) \label{fourier-zernike-coef}
\end{equation}
where $f_{lmn}$ are the spectral coefficients, $\mathcal{Z}_l^{m}(\rho, \theta)$ is the Zernike polynomials, and $\mathcal{F}^n(\zeta)$ is the Fourier components corresponding to toroidal mode numbers. The maximum resolution chosen for numerical implementation is described by $L, M$ and $N$ for radial, poloidal and toroidal directions, respectively. Zernike polynomials can be described as follows,
\begin{equation}
    \mathcal{Z}_l^{m}(\rho,\theta) = \begin{cases}\begin{aligned}
            &\mathcal{R}_l^{|m|}(\rho) \cos(|m|\theta) & \text{for } m\geq 0 \text{ and } 0\leq\rho\leq 1 \\[.3cm]
            &\mathcal{R}_l^{|m|}(\rho) \sin(|m|\theta) & \text{for } m < 0  \text{ and } 0\leq\rho\leq 1
        \end{aligned}
    \end{cases}
\end{equation}
where $\mathcal{R}_l^{|m|}(\rho)$ is called the radial part of the Zernike polynomials and is defined as,
\begin{equation}
    \mathcal{R}_l^{|m|} (\rho) = \mathlarger{\mathlarger{\sum}}_{s=0}^{(l-|m|)/2} (-1)^s \binom{l-s}{s} \binom{l-2s}{\frac{l-|m|}{2}-s}\hspace{0.1cm} \rho^{l-2s} \label{radial-part-eq-binom}
\end{equation}
where the mode numbers $l$ and $m$ are defined such that $l\geq0$, $|m|\leq l$, and the difference $l-|m|$ must be even \citep{zernikeBeugungstheorieSchneidenverfahrensUnd1934, zernikeDiffractionTheoryKnifeEdge1934}. For the LCFS where $\rho=1$, the radial part of the Zernike polynomials is always equal to 1. This results in a simple linear relation between the LCFS boundary and global coefficients,
\begin{equation}
    R^b_{mn} = \sum_l R_{lmn} \hspace{2cm} Z^b_{mn} = \sum_l Z_{lmn} \label{lcfs_rel}
\end{equation}
where $R^b_{mn}$ and $Z^b_{mn}$ are the LCFS boundary coefficients, which are either given by the user or optimized for some objective.

The way equilibrium is solved in \texttt{DESC} can be summarized by the following optimization problem,
\begin{equation}
    \begin{split}
        \min_{x} ||\mathbf{J}(x)\times\mathbf{B}(x)-\nabla p(x)||_2^2 \\
        \text{s.t. } \mathbf{A}x = b 
    \end{split}
    \label{opt-eq-solve}
\end{equation}
where $\mathbf{A}$ is the linear constraint matrix composed of Equation \ref{lcfs_rel} along with other constraints (such as fixed profiles, gauge freedom of $\lambda$ etc.) and $x$ is the state vector defining the equilibrium (i.e. $R_{lmn}, Z_{lmn}, \lambda_{lmn}$, coefficients of the profiles, boundary, and the axis). The linear constraint can then be eliminated by writing $x = \mathbf{N}y + x_p$, where $\mathbf{N}$ is a basis for the nullspace of $\mathbf{A}$ so that $\mathbf{AN}=0$, $y$ is the reduced state vector, and $x_p$ is a particular solution of $\mathbf{A}x_p=b$.

To adapt the equilibrium solution to the Poincaré boundary condition, we expand the prescribed cross-section in the same basis restricted to that plane, with coefficients $R^p_{lm}$ and $Z^p_{lm}$, and evaluate Equation \ref{fourier-zernike-coef} at $\zeta=0$,
\begin{equation}
    \sum_{l, m}^{}R^p_{lm} \mathcal{Z}_l^{m} (\rho, \theta) =  \sum_{l, m}^{} \left( \sum_{n}^{} \mathcal{F}^n(\zeta=0) R_{lmn}  \right)  \mathcal{Z}_l^{m} (\rho, \theta)
\end{equation}
Since the Zernike polynomials and the Fourier series are linearly independent, we have,
\begin{equation}
    R^p_{lm} = \sum_{n}^{} R_{lmn} \mathcal{F}^n(\zeta=0)  \label{poincare-rel}
\end{equation}
and,
\begin{equation}
    \mathcal{F}^n(\zeta=0) = \begin{cases}
        0 & \text{for } n<0\\
        1 & \text{for } n\geq0\\
    \end{cases} \label{zeros-in-fourier}
\end{equation}

The same relations hold for $Z$ and for the poloidal stream function $\lambda$, and the constraint on $\lambda$ can be imposed or omitted independently of the geometric one. Therefore, the implementation of the new boundary condition only changes the linear constraint matrix $\mathbf{A}$ in Equation \ref{opt-eq-solve}, hence most of the machinery in \texttt{DESC} can be kept unchanged.

Nothing in this construction requires the plane to be at $\zeta=0$. Evaluating Equation \ref{fourier-zernike-coef} at an arbitrary $\zeta_0$ gives a relation of the same form, with $\mathcal{F}^n(\zeta_0)$ in place of $\mathcal{F}^n(\zeta=0)$, so any toroidal angle can be used, or multiple of them can be imposed. We choose $\zeta=0$ because it is a symmetry plane of a stellarator-symmetric configuration, as is $\zeta=\pi/N_{fp}$. On either plane the sine modes drop out of the constraint and the cross-section is up-down symmetric, so it can be represented in the same symmetric basis as the equilibrium, whereas a cross-section at any other angle would require both the sine and cosine modes even when the equilibrium itself is stellarator symmetric.

It is worth noting that, due to the relation in Equation \ref{zeros-in-fourier}, evaluating at $\zeta=0$ removes the sine modes from the constraint entirely and collapses the cosine modes into a single sum for each $(l,m)$. The Poincaré boundary condition therefore fixes one linear combination of the toroidal coefficients belonging to a given $(l,m)$ and leaves the rest free. Counting the rows of each, which is done in Appendix \ref{app:counting}, shows this to be the case whenever $N \gtrsim (2L-M)/8$, or $N \gtrsim M/8$ for the common choice $L=M$. For $L=M=12$ and $N=6$, the resolution used for the quasi-helical optimization of \S\ref{sec:stage1}, the Poincaré boundary condition imposes 49 constraints on the 595 spectral coefficients of $R$, whereas the LCFS imposes 163. The constraint matrix consequently has fewer rows than its LCFS counterpart and a correspondingly larger nullspace, and hence more free parameters. What matters as much as the number of free parameters is where they lie: the freed coefficients are the ones that carry the toroidal variation of the surfaces, which a prescribed LCFS holds fixed at every toroidal angle. The heliotron of \S\ref{sec:eqsolve}, solved at a high radial and a low toroidal resolution, is constrained by more rows in total than its fixed-LCFS counterpart, and still converges to a lower force error due to the freedom in the toroidal direction.

This counting describes the space in which the solver searches, and not the set of equilibria contained in it. An equilibrium is defined by $\mathbf{J}\times\mathbf{B}=\nabla p$, and Equation \ref{opt-eq-solve} is a non-convex minimization used to satisfy that equation approximately: a converged state is an approximate equilibrium when its volume-averaged normalized force error is small enough, which is usually taken to mean around $1\%$ or below\footnote{What counts as small enough is not settled in the stellarator community, and the checks applied in practice are external ones, such as comparing against a higher fidelity calculation that does not assume nested surfaces, or tracing field lines of the field obtained from a coil set for vacuum equilibria. A criterion internal to the nested-surface solver would be preferable, but establishing one is beyond the scope of this paper.}, and a local minimum with a large residual is only a point at which the solver stalls. The larger nullspace of the Poincaré boundary condition gives the solver more directions in which to move, so more of these minima are within reach, and in practice several of them can attain a force error as low as, or lower than, the fixed-LCFS solution. Two such states can have different LCFS shapes and volumes even though they share the prescribed cross-section, so which one is found depends on the initial guess. We treat this as a means of exploring the design space, made available by the additional freedom of the Poincaré boundary condition.

As stated in the introduction, this choice is algorithmic rather than analytical. Neither boundary condition guarantees an island-free field, and fixing a cross-section says nothing about whether nested surfaces exist. In either case the solver returns the best nested representation available at the given resolution. If relaxing the constraint lowers the force error, one might ask why we keep a boundary condition at all, since removing it entirely and leaving every spectral coefficient free would lower the force error further. The reason is that the boundary condition is also what tells the solver which device we are looking for. The prescribed cross-section still sets the size and shape of the plasma in that plane, such as its minor radius and elongation, while the geometry away from the plane is left free.

\section{Results}
\subsection{Equilibrium Solve}\label{sec:eqsolve}
As described in the previous section, one of the inputs to an equilibrium solve is the boundary shape. One must specify the initial nested coordinate mapping in the rest of the volume from which to then proceed with the equilibrium solve, which in \texttt{DESC} means specifying the initial Fourier-Zernike coefficients for $R,Z$, and $\lambda$. When the boundary is the LCFS, the poloidal stream function $\lambda$ is set to 0 and via Equation \ref{lcfs_rel} some of the $R$ and $Z$ coefficients are defined, and the rest can be initialized by scaling down that surface around an initial axis, which for convenience can be taken to lie midway between the $\theta=0$ and $\theta=\pi$ curves of the boundary, or by more sophisticated initialization methods discussed in the literature \citep{tecchiolli_constructing_2024, babin_construction_2025}. 

\begin{figure}
    \centering
    \includegraphics[width=\linewidth]{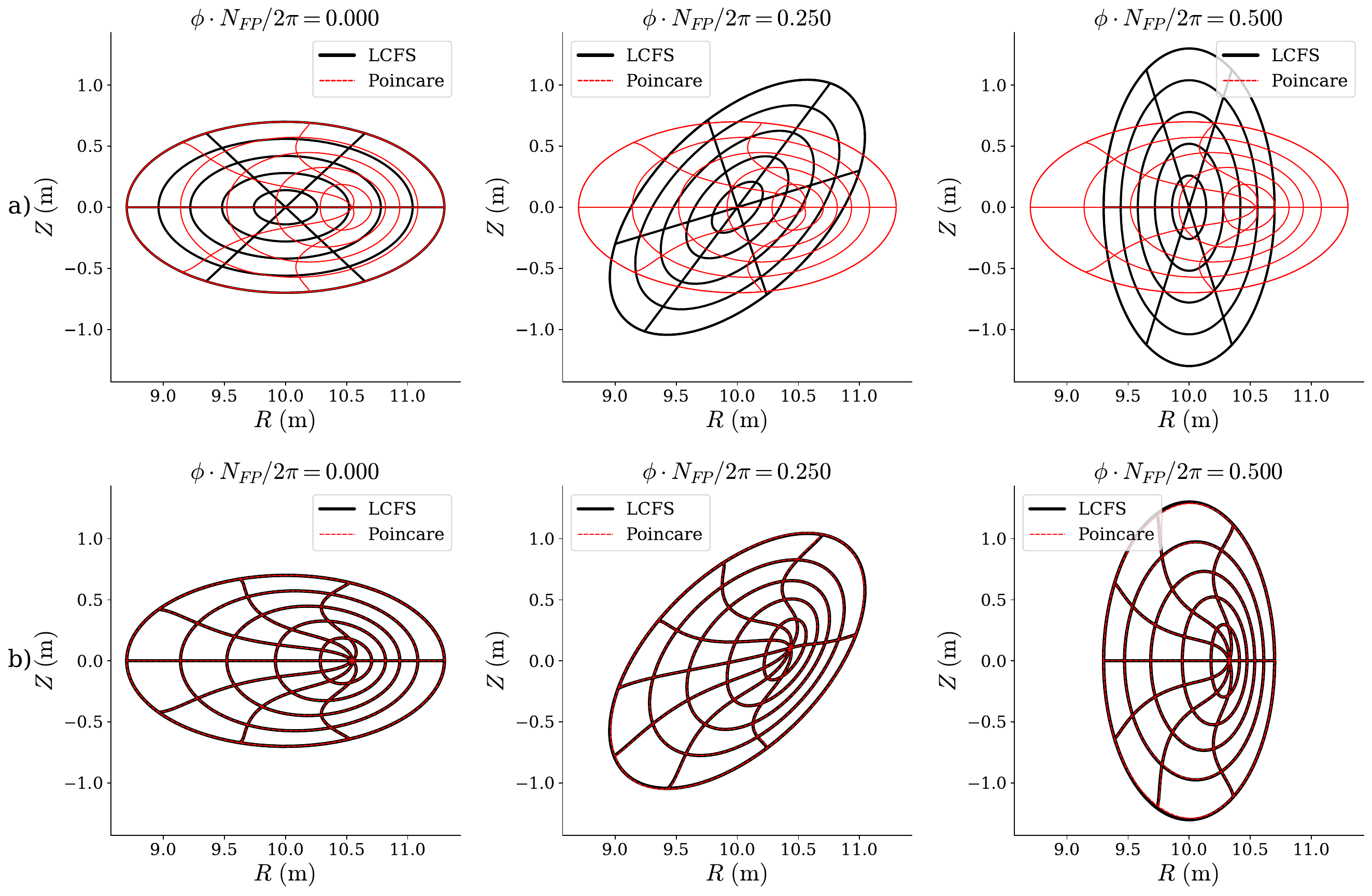}
    \caption{a) The initial and b) final flux surface shapes for LCFS (red) and Poincaré boundary conditions (black) of a heliotron device with $N_{fp}=19$. The lines that cross the constant $\rho$ surfaces are the constant straight field line angle $\vartheta = \theta+\lambda$.}
    \label{fig-initfinal}
\end{figure}

For the Poincaré boundary condition, only the cross-section at $\zeta=0$ is known, so the natural initial guess is to revolve it toroidally and start from an axisymmetric shape with $N=0$. Here $\lambda$ is initialized in the same way, from the $\lambda$ of the cross-section. Independent of the initial guess, the $\lambda$ cross-section can be fixed for comparison purposes or left free during the solve\footnote{When $\lambda$ is free, only the geometry of the flux surface shapes is given, whereas the inclusion of $\lambda$ also provides information on the straight field line angle $\vartheta$ of the solved equilibrium on the $\zeta=0$ cross-section.}. The cross-section used here is taken from the $\zeta=0$ cut of the original fixed LCFS solved equilibrium, so that the two solutions can be compared. The initial guess surfaces for a heliotron device with $N_{fp}=19$ and $\beta=10\%$, solved at a resolution of $L=24$, $M=12$ and $N=3$, are shown in Figure \ref{fig-initfinal} for both LCFS and Poincaré boundary conditions with fixed $\lambda$. The pressure and rotational transform profiles are $p(\rho) = 1.8\times10^{4}(\rho^2 - 1)^2$ Pa and $\iota(\rho) = 1 + 1.5\rho^2$, and the total enclosed toroidal flux is $\Psi = 1$ Wb.

\begin{figure}
    \centering
    \includegraphics[width=\linewidth]{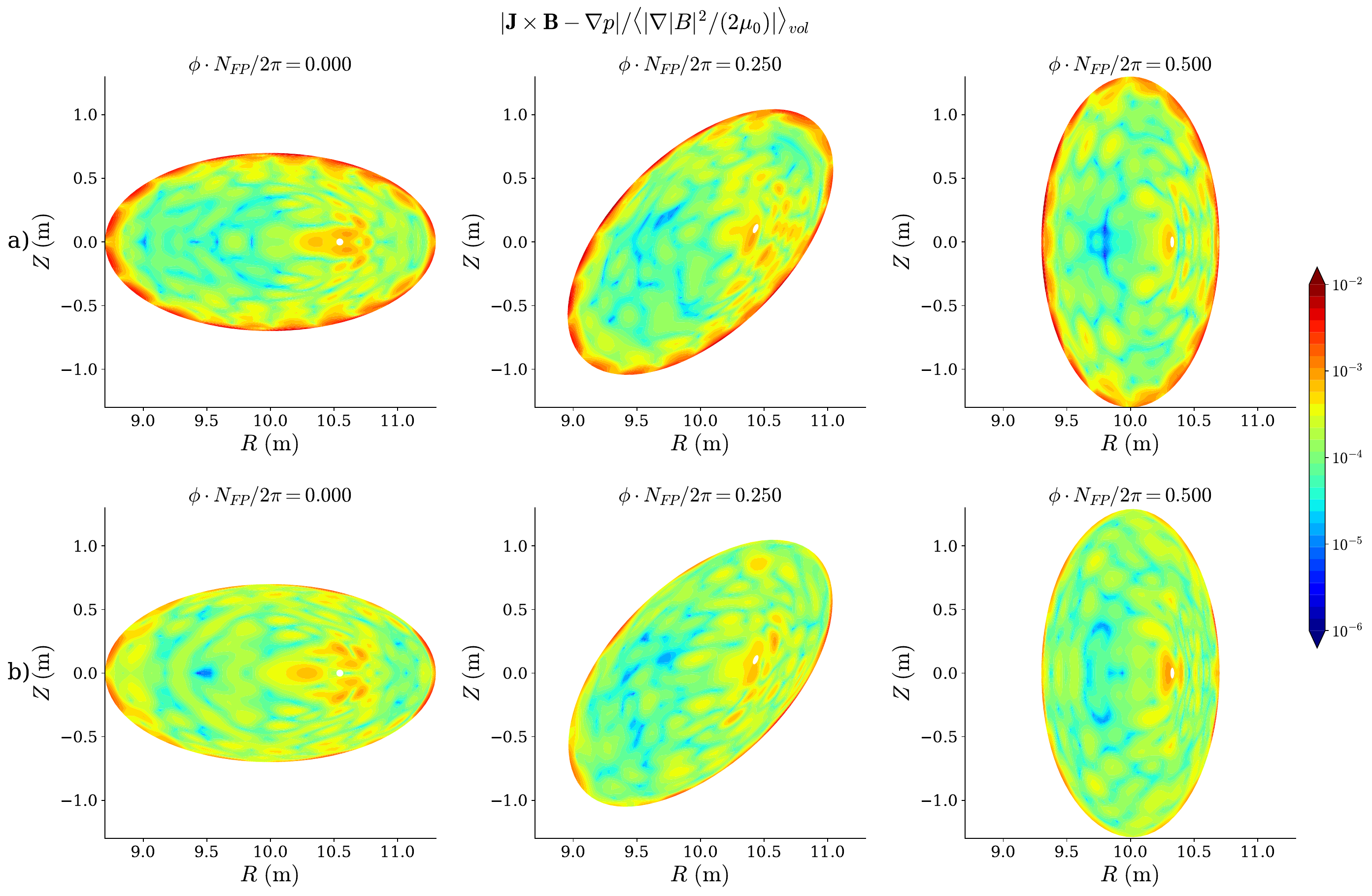}
    \caption{The ideal MHD force error normalized by volume-averaged magnetic pressure for a) the LCFS solution and b) the Poincaré solution of the HELIOTRON device with $N_{fp}=19$. The errors are plotted in log scale for 3 equispaced toroidal angles between 0 and $\pi/N_{fp}$ , over the range $10^{-2}$ to $10^{-6}$.}
    \label{force-helio}
\end{figure}

The equilibrium solve is conducted in stages, increasing the toroidal resolution $N$ before each step until the resolution of the original equilibrium is reached while keeping the $L$ and $M$ resolution the same. In Figure \ref{fig-initfinal}, it can be seen that the final surface shapes are visually similar. Moreover, the obtained equilibrium has good force balance. In Figure \ref{force-helio}, we observe that in the LCFS solution, the error is accumulated around the LCFS boundary, whereas in the Poincaré solution, this trend is observed more around the fixed cross-section. The final volume-averaged normalized force error is 3E-4 and 1.7E-4 for LCFS and Poincaré, respectively.

Although the final equilibria are very similar for the heliotron device with $N_{fp}=19$, we see that in other cases where the LCFS boundary has more shaping, the axisymmetric initial guess tends to give different equilibria, and in many cases the final equilibrium expands well beyond the original device. Part of this behaviour originates in the objective function rather than in the boundary condition itself. With the LCFS prescribed, the enclosed cross-sectional area, $A$, is fixed, and at fixed toroidal flux $\Psi$ so is the field strength, since $B \sim \Psi/A$; expansion is simply not available to the solver, and the force error can therefore be normalized once at the start of the solve, as is done in \texttt{DESC}. Under the Poincaré boundary condition the equilibrium is free to expand away from the prescribed plane, resulting in a weaker field, and because $||\mathbf{J}\times\mathbf{B}||$ scales as $B^2$, an inflated equilibrium computes a smaller residual against a normalization that no longer reflects its field strength. Normalizing by a quantity recomputed at each iterate, such as the volume-averaged $B^2$, could help here, since a measure of that kind is unchanged by a uniform weakening of the field. For the same reason, every force error quoted in this paper is normalized by the volume-averaged magnetic pressure of the equilibrium it refers to, so that the comparisons that follow are not affected by differences in size or field strength between solutions. The volume and aspect ratio that a prescribed LCFS carries implicitly can likewise be supplied explicitly as additional objectives; however, this makes it a multi-objective optimization which does not guarantee the minimum of the force balance error.

The gap in the number of free parameters also widens as $N$ grows relative to $L$, since the LCFS constrains one coefficient for every $(m,n)$ pair whereas the cross-section constrains one for every $(l,m)$ pair, and a high toroidal resolution is usually needed for low $N_{fp}$ devices. Combined with an initial guess that is far from the intended configuration, this leads the solver to a different local minimum, where the additional freedom generally yields better force balance with a very different LCFS shape. Given the higher freedom, the Poincaré constraint equilibrium solution depends strongly on the initial guess. In the following section we show that beginning from an existing LCFS solution keeps the solution nearby while still improving the force balance.

\subsection{Reduced MHD Force Error}
\begin{figure}
    \centering
    \includegraphics[width=\linewidth]{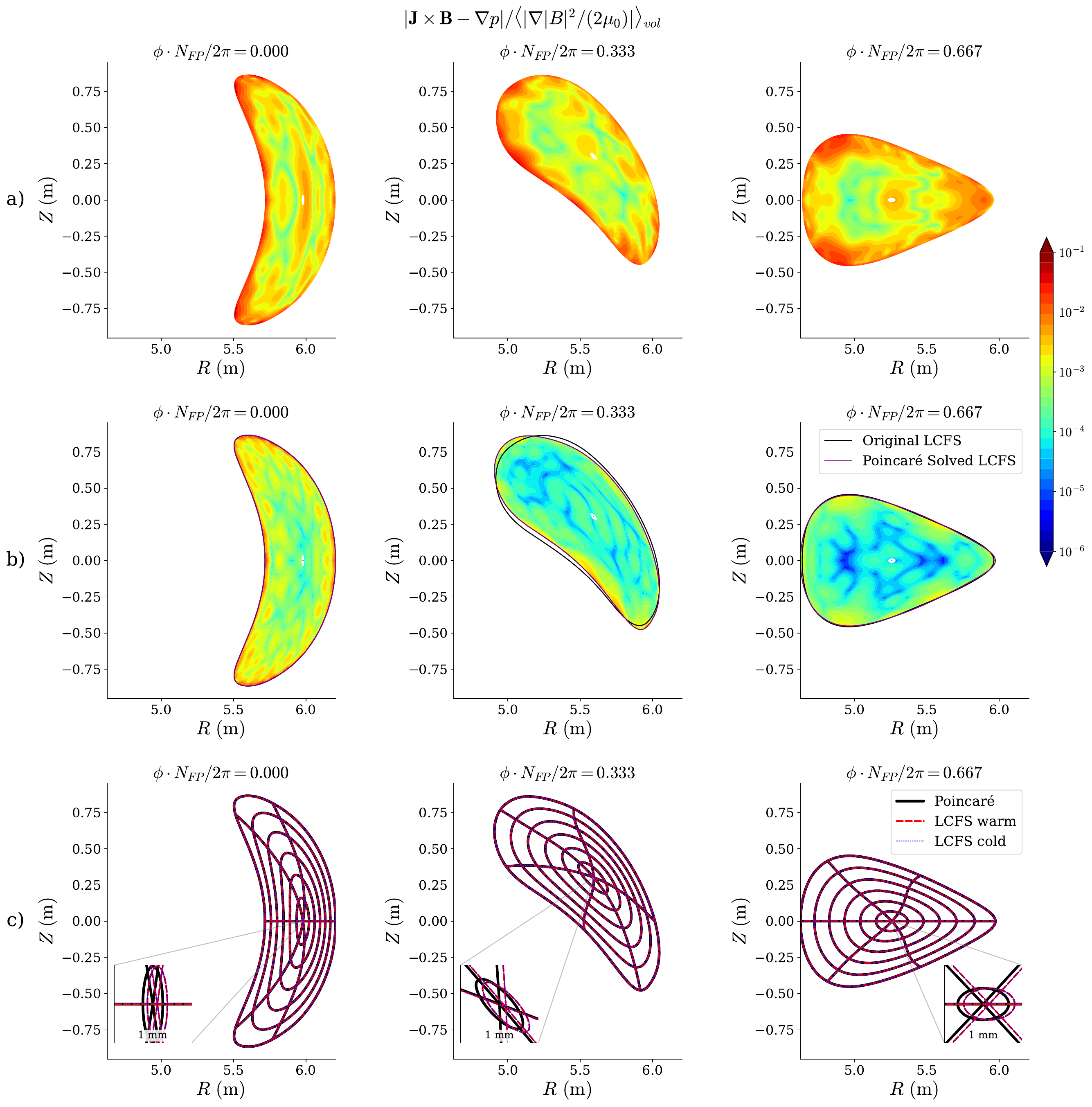}
    \caption{The ideal MHD force error normalized by volume-averaged magnetic pressure (a) for the LCFS solution and (b) for the Poincaré solution of the W7-X like device. All solves use ftol=1E-3, which is the relative decrease in the objective value between two consecutive iterations, and everything is plotted for 3 equispaced toroidal angles between 0 and $\pi/N_{fp}$. For comparison, part (b) additionally shows both the original LCFS of part (a) and the modified LCFS after the Poincaré solve. Part (c) compares the flux surfaces and $\vartheta$ contours of the Poincaré solution with those of the traditional fixed-LCFS solve of the LCFS boundary obtained from the Poincaré solution, started either from the Poincaré solution as the initial guess (warm) or from the standard scaled-boundary initial guess (cold). The insets are 1 mm wide and zoom in on the magnetic axis, showing the $\rho=5\times10^{-4}$ surface of each solution.}
    \label{force-w7x}
\end{figure}

In the previous section, we noted that using the straightforward axisymmetric initial guess for the Poincaré boundary condition usually results in a different final equilibrium than the original LCFS solution. However, one can expect to overcome this by using the LCFS solution as the initial guess. The aim of this approach is to improve the relatively high force error for the fixed-LCFS solution observed in Figure \ref{force-helio} and Figure \ref{force-w7x} around the LCFS. Since the Poincaré and LCFS boundary conditions act on different $R_{lmn}$ and $Z_{lmn}$ coefficients, this helps the optimizer to discover the previously forbidden regions of the optimization space while still being close to the LCFS solution. We therefore restrict the problem to a single question: whether an equilibrium that stays close to that solution, but has a lower force error, can be obtained from the Poincaré boundary condition alone, with no additional objectives imposed. For this section and later ones, we won't fix the poloidal stream function $\lambda$, in other words, we are just imposing a geometric constraint on flux surface shapes at $\zeta=0$ toroidal cut without giving information on the straight field-line angle.

In Figure \ref{force-w7x}, we take a W7-X-like equilibrium at $\beta=2\%$, average minor radius $a=50.6$ cm, and a resolution of $L=M=N=12$, which is solved for the force balance while keeping the $\zeta=0$ cross-section of the previous solution. Its pressure profile is $p(\rho) = 1.856\times10^{5}(\rho^2 - 1)^2$ Pa, the rotational transform increases monotonically from 0.856 on axis to 0.963 at the edge, and the total enclosed toroidal flux is $\Psi = 2.133$ Wb. It can be seen that solving the equilibrium by imposing Poincaré boundary conditions improves the force balance substantially, with the volume-averaged force error normalized by volume-averaged magnetic pressure dropping from 2.86E-3 to 2.70E-4. The average LCFS displacement is 1.9$\%$ of the mean minor radius, with a maximum shift of 7.4$\%$, while the plasma volume changes by just 0.6$\%$. The cross-section that is fixed sees only minor improvement, which is expected due to the limited freedom there.

We then check that this solution is an equilibrium in the traditional sense as well. We take its LCFS and solve the fixed-LCFS problem again, using the Poincaré solution as the initial guess. We refer to this as the warm start in Figure \ref{force-w7x}(c). Even though the $\zeta=0$ cross-section is not constrained in this solve, so the internal surfaces are free to move away from it, the converged equilibrium stays very close to the one it started from, with a force error of 2.09E-4 and a magnetic axis that moves by at most 0.019$\%$ of the average minor radius. Since that solve begins at the very state it is meant to test, we also repeat it from the standard scaled-boundary initial guess, so that the solver is given only the LCFS boundary and has no knowledge of the Poincaré solution. We refer to this as the cold start in Figure \ref{force-w7x}(c). It converges to the same equilibrium as the previous warm start, with the same force error of 2.09E-4 and the magnetic axis again lying within 0.019$\%$ of the minor radius of the Poincaré one. The flux surfaces of the three solutions are indistinguishable in Figure \ref{force-w7x}(c), where the insets resolve the axis region down to 1 mm. The traditional fixed-boundary formulation therefore reaches this equilibrium on its own.

Admittedly, this procedure modifies the equilibrium, which may have been previously optimized for specific physics metrics; this shift potentially degrades those optimized values. Nevertheless, the approach remains valuable as it demonstrates that solving the equilibrium with Poincaré boundary conditions generally yields a superior force balance, which many physics metrics needs to be computed correctly (i.e. Mercier stability comparison for stellarators is shown in \citep{paniciDESCStellaratorCode2023} and the effect of Grad-Shafranov error for stability and transport analysis is discussed in \citep{xing_cake_2021, jiang_kinetic_2021}). In the following section, we will demonstrate how Poincaré variables can be integrated directly into the equilibrium optimization process, traditionally called the stage-one optimization, to maintain high-fidelity force balance throughout the entire procedure.

\subsection{Stage-One Optimization}\label{sec:stage1}
This section presents how \texttt{DESC} performs equilibrium optimization, as described in \citep{dudtDESCStellaratorCode2023}, and how we modified it to use Poincaré variables. A general stage-one optimization problem can be written as follows:
\begin{equation}
    \begin{split}
        \min_{c} \hspace{1cm} &f(\textbf{x, c}) \\
        \text{s.t. } \hspace{1cm} &g(\textbf{x, c}) = 0\\
        &\mathbf{Ac} = \mathbf{b}
    \end{split}
\end{equation}
where $x$ includes the 3D spectral coefficients of the equilibrium, $c$ is the optimizable parameters of the boundary and profiles from which we solve the equilibrium, $f$ is the objective to minimize, $g$ is the nonlinear ideal MHD equilibrium constraint, and $\mathbf{A}$ is the linear constraint matrix which is used during the continuation method to selectively fix certain high-order boundary modes and also to fix the profiles and total toroidal magnetic flux. To deal with the non-linear constraint, \texttt{DESC} solves the equilibrium with the current $c$ values to find the projected $x$ values on to this non-linear constraint. Once the projected state is obtained, a gradient-based optimization can be conducted. We first expand the $f$ and $g$ using a first order Taylor series,
\begin{align}
    f(x + \dx,c + \dc) = f(x, c) + \frac{\partial f}{\partial x} \dx + \frac{\partial f}{\partial c} \dc \label{f_taylor}\\
    g(x + \dx,c + \dc) = g(x, c) + \frac{\partial g}{\partial x} \dx + \frac{\partial g}{\partial c} \dc \label{g_taylor}
\end{align}
We aim to find the step $\dc$ that minimizes the predicted cost $f(x + \dx,c + \dc)$. One can remove the explicit dependence on $\dx$ by assuming that $g(x + \dx,c + \dc) = g(x, c) \sim 0$ by solving the equilibrium before and after, which holds better for Poincaré solve. This results in the following relation for each optimization step,
\begin{equation}
    f(x + \dx,c + \dc) = f(x, c) + \left[ \frac{\partial f}{\partial c} - \frac{\partial f}{\partial x} \left( \frac{\partial g}{\partial x} \right) ^{\dagger} \frac{\partial g}{\partial c} \right] \dc \label{g_linear}
\end{equation}

The term in the square brackets is the Jacobian of the objective at fixed force error with respect to the optimization variables, and $\dagger$ is the pseudo-inverse. Each optimization step tries to minimize $f(x + \dx,c + \dc)$, and this can be done in many different ways. The algorithm we use for the remaining of the optimizations is a trust-region based solver. To make this procedure work when the optimization degrees of freedom $c$ are the Poincaré variables, we update all the partial derivatives as well as replace the internal equilibrium solver to the fixed Poincaré section variant we discussed above.

Now, we will show an example optimization for a vacuum quasihelical symmetric (QH) stellarator. The exact problem is as follows,
\begin{equation}
    \begin{split}
        \min_{c=(R^p_{lm}, Z^p_{lm})} w_{qs}f_{qs}^2 + w_{ar}f_{ar}^2 + w_{B}f_{B}^2 \\
        \text{s.t. } \hspace{1cm} g(x, c)=0 \hspace{0.5cm}  Ac=b
    \end{split}
\end{equation}
where $c=(R^p_{lm}, Z^p_{lm})$ are the coefficients defining the Poincaré cross-section. The problem setup is very similar to the one employed in \cite{landremanMagneticFieldsPrecise2022}. The objective function terms include $f_{qs}$, the two-term quasi-symmetry error \citep{helanderIntrinsicAmbipolarityRotation2008a, rodriguezMeasuresQuasisymmetryStellarators2022a}, and $f_{ar}, f_{B}$, which are the residuals from the target aspect ratio (AR) and average magnetic field, respectively. The terms $w_{qs}, w_{ar}, w_{B}$ are the corresponding weights for each objective. For the results presented here, we used a continuation in Zernike coefficients $R^p_{lm}$ and $Z^p_{lm}$ where at each step higher $l$ (and corresponding $m$) coefficients are allowed to change, and we also included exponential spectral scaling for the optimization parameters with $\alpha=1.2$ and $\max(|l|, |m|)$ (the $L_\infty$ norm) scaling pattern \citep{jangExponentialSpectralScaling2026}.

The two-term QS metric is given by,
\begin{equation}
f_{qs}= \frac{(M\iota - N)(\bm{B}\times\nabla\psi)\cdot\nabla B - (MG+\iota I)\bm{B}\cdot\nabla B}{\langle|\bm{B}|\rangle^3}
\label{f_qs}
\end{equation}
where $\bm{B}$ is the magnetic field, $\nabla B$ is the gradient of the magnitude of the magnetic field, $(M,N)$ are the QS helicity with $(M,N)=(1,4)$ is chosen for the QH case, $\psi$ is the normalized toroidal flux, $I,G$ are the Boozer toroidal and poloidal currents and $\langle|\bm{B}|\rangle$ is the flux-surface average magnetic field magnitude\footnote{It is worth noting that we had to normalize the standard two-term QS metric defined in \texttt{DESC} by the cube of the flux-surface average magnetic field to prevent the optimizer from cheating by reducing the magnetic field strength. This performed much better than tuning the weight for the average magnetic field strength. The importance of the normalization term has also been mentioned in \cite{landremanMagneticFieldsPrecise2022}, however, in practice, LCFS-based optimizations seem to be less dependent on the inclusion of this term.}, respectively. During the optimization, we compute $f_{qs}$ using the local field at a grid of points on $\rho=[0.6, 0.8, 1.0]$ surfaces and minimize it with the least-squares approach.

Another important distinction from the LCFS optimization is that for the Poincaré variant, we included an additional volume average magnetic field strength objective, which is mostly enforced for the LCFS version by the target aspect ratio, the fixed $R^b_{00}$ mode of the LCFS and the fixed toroidal magnetic flux, $\Psi$. We have seen that without such a penalty, the optimizer sometimes finds a similar solution that has a larger major/minor radius while keeping the aspect ratio the same. Similarly, a volume objective can be added as regularisation for the problem. For this particular example, we observe that the optimizer can find good solutions without needing it explicitly. On a similar point, we didn't fix $R^p_{00}$ since it only constrains the average radius of the Poincaré cross-section without giving any other geometric meaning as $R^b_{00}$. The inclusion of a fixed $R^p_{00}$ constraint is found to deteriorate the optimization performance significantly. Apart from these differences, the optimization is conducted in the same way with similar initial guess, which has an LCFS of near circular cross sections with axis torsion\footnote{More specifically, the initial guess used for the LCFS optimization in \cite{landremanMagneticFieldsPrecise2022} is solved by fixed Poincaré boundary condition, and this result is set as the initial guess for the optimization. The extra Poincaré solve didn't change the equilibrium in a distinguishable way.}.

\begin{figure}
    \centering
    \includegraphics[width=\linewidth]{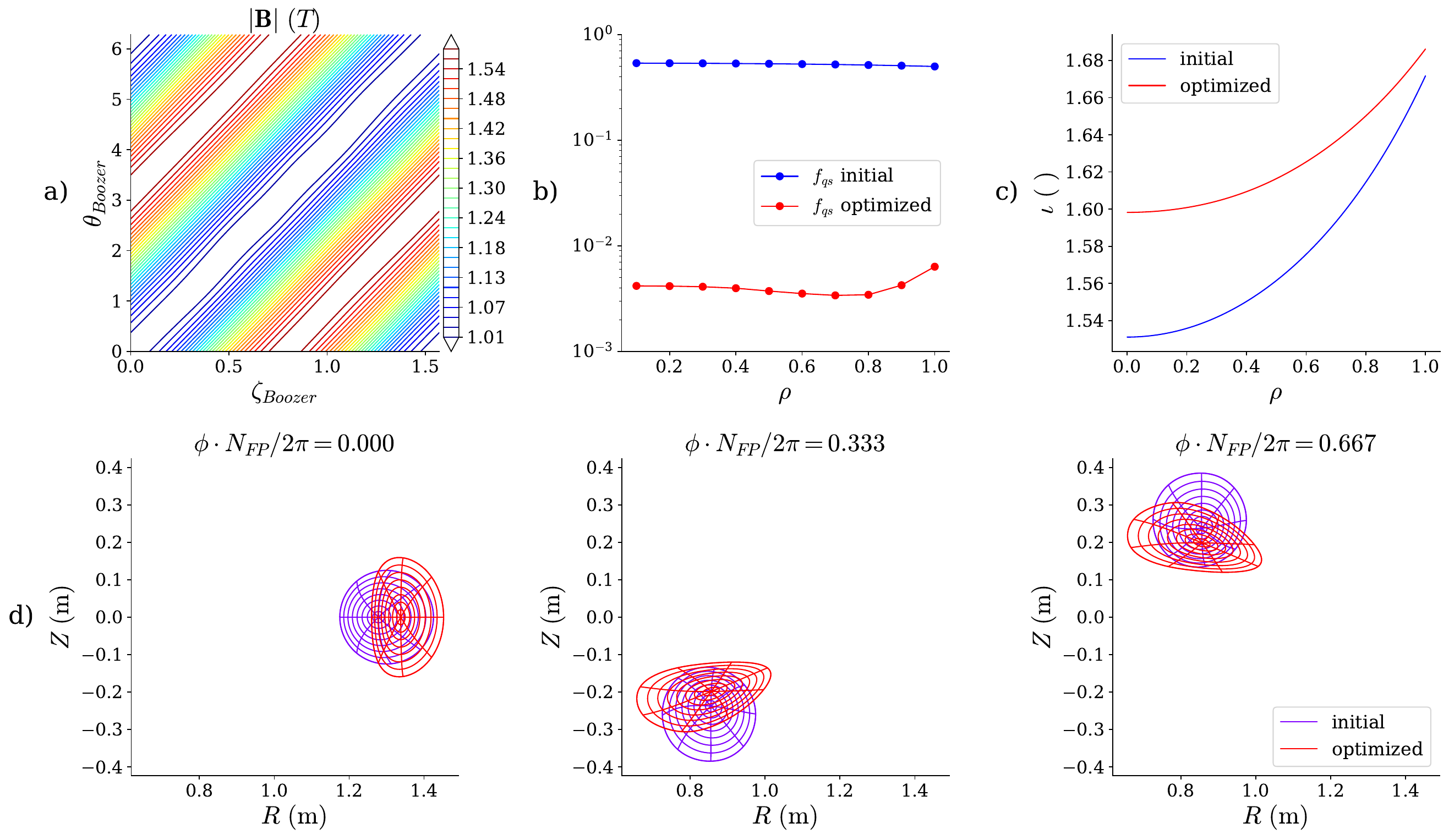}
    \caption{a) $|B|$ level sets in Boozer coordinates for $\rho=1$ surface, b) Quasisymmetry errors before and after optimization, c) Rotational Transform of the initial and optimized equilibria, d) The flux surfaces of the initial and optimized equilibrium.}
    \label{qh-results}
\end{figure}

The QH optimization results shown in Figure \ref{qh-results} are obtained with the following weights and targets,
\begin{equation}
    \begin{aligned}
        \langle|\bm{B}|\rangle = 1.26 \text{T} \hspace{0.5cm}& w_B = 1 \\
        AR = 8 \hspace{0.5cm}& w_{ar} = 3 \\
        & w_{qs} = 3 / |f_{qs0}| 
    \end{aligned}
\end{equation}
where $|f_{qs0}|$ is the initial value of the QS objective without any weights which is 33.3 for this initial guess, and it has helped to reduce the scanned weight combinations. The volume of the initial guess and the final optimized solution are both 0.307 $\text{m}^3$. The number of field period symmetry is chosen to be $N_{fp}=4$.

The quasi-symmetry error falls by roughly two orders of magnitude, with $f_{qs}$ dropping from 0.499 to 6.3E-3 at the boundary and by a somewhat larger factor further in, while the aspect ratio and the volume-averaged field strength stay at their targets. The volume-averaged force error normalized by volume-averaged magnetic pressure remains between 1.5E-5 and 2.5E-5 at every step of the continuation, so the quasi-symmetry is always evaluated on an equilibrium that is well converged in force balance. A further comparison of the optimized results obtained with traditional LCFS optimization is included in Appendix \ref{app:comparison}.

\section{Conclusion}
In this work, the Poincaré cross-section is proposed as an alternative to the LCFS boundary condition for three-dimensional ideal MHD equilibria, and implemented in \texttt{DESC}. Because the condition enters only through the linear constraint matrix $\mathbf{A}$, the rest of the solver and of the optimization machinery is very similar. Prescribing the geometry on a single toroidal plane fixes one linear combination of the toroidal coefficients for each $(l,m)$ and leaves a considerably larger nullspace than a fixed LCFS does, and it is this additional freedom that the solver exploits. We show that solving the equilibrium with the new boundary condition gives lower force errors in a variety of cases. Moreover, the same coefficients can serve directly as design variables, and we used them to obtain a quasi-helical configuration while maintaining force balance to high fidelity throughout the optimization which is crucial for computing the physics objectives reliably.

A few points should be kept in view. The Poincaré cross-section is an algorithmic device rather than a statement about the actual field: it does not guarantee island-free equilibria, and being the less restrictive of the two conditions, it makes the local minima of the force-error minimization easier to encounter, which is why a solve started far from the intended configuration can converge to one of very different character. Starting instead from an existing LCFS solution keeps the solver in the intended region, and is the mode in which we find the method most useful.

The two boundary conditions also constrain different things, so the objectives need small changes. A prescribed LCFS fixes the plasma volume and the aspect ratio, and with a fixed toroidal flux it fixes the field strength as well. A prescribed cross-section fixes none of these away from its own plane, and the optimizer will use any freedom the objectives leave it, as any optimizer would. Normalizing the residuals by quantities computed at the current iterate, and adding objectives for what a prescribed LCFS gives for free, is enough to keep the problem well posed. Set up in this way, the Poincaré boundary condition offers an inexpensive way to reduce the force error that most physics metrics used in stellarator design implicitly assume to vanish.

\section{Acknowledgments}
This work is funded through the SciDAC program by the US Department of Energy, Office of Fusion Energy Science and Office of Advanced Scientific Computing Research under contract No. DE-AC02-09CH11466, as well as DE-SC0022005 and the Hidden Symmetries grant from the Simons Foundation (560651). The United States Government retains a non-exclusive, paid-up, irrevocable, world-wide license to publish or reproduce the published form of this manuscript, or allow others to do so, for United States Government purposes.

\section{Data availability statement}
The source code to generate the results and plots in this study is openly available on GitHub at \url{https://github.com/PlasmaControl/DESC} under the \texttt{poincare-bc-paper} branch and \url{https://github.com/PlasmaControl/Poincare-Paper}. The full details of the equilibria used here, including the profile coefficients and boundary shapes, can be found in those repositories.

\section{Declaration of generative AI and AI-assisted technologies in the writing process}
During the preparation of this work, the author Yigit Gunsur Elmacioglu used Anthropic's Claude Opus 5 in order to increase the readability of the text. After using this tool, the authors reviewed and edited the content as needed and take full responsibility for the content of the publication.

\appendix
\section{Counting the linear constraints}\label{app:counting}
Each boundary condition contributes one row to the linear constraint matrix $\mathbf{A}$ of Equation \ref{opt-eq-solve} for every mode of the geometry it prescribes. We denote by $N_p$ the number of rows contributed by the Poincaré boundary condition, and by $N_b$ the number contributed by the prescribed LCFS, with a superscript $R$, $Z$ or $\lambda$ when we refer to the rows acting on one of these functions. Throughout, we use the ANSI Zernike indexing with $L\geq M$, both even; other choices change the arithmetic but not the conclusions. It is convenient to count the Zernike pairs $(l,m)$ of the basis separately for the two signs of $m$,
\begin{equation}
    P = \frac{2LM + 2L + 2M + 4 - M^2}{4} \hspace{2cm} Q = \frac{M(2L + 2 - M)}{4} \label{PQ-counts}
\end{equation}
where $P$ counts the pairs with $m\geq0$, whose poloidal dependence is $\cos(m\theta)$, and $Q$ those with $m<0$, whose poloidal dependence is $\sin(|m|\theta)$. Both follow from summing $\lfloor (L-|m|)/2 \rfloor + 1$, the number of radial modes available to a given $m$, over the relevant range of $m$, and both reduce to $P=(M+2)^2/4$ and $Q=M(M+2)/4$ for the common choice $L=M$.

Without stellarator symmetry, each of $R$, $Z$ and $\lambda$ carries all of these pairs combined with all $2N+1$ toroidal modes, that is $(P+Q)(2N+1)$ coefficients. The Poincaré condition, Equation \ref{poincare-rel}, gives one relation for every $(l,m)$ pair, since at $\zeta=0$ the $\sin(nN_{fp}\zeta)$ modes vanish while the $\cos(nN_{fp}\zeta)$ modes collapse into a single sum. The LCFS condition, Equation \ref{lcfs_rel}, gives one relation for every $(m,n)$ pair of the boundary surface. Per function,
\begin{equation}
    N_p = P + Q = \frac{(M+1)(M+2)}{2} \hspace{2cm} N_b = (2M+1)(2N+1)
\end{equation}

Stellarator symmetry, $R(\rho,-\theta,-\zeta)=R(\rho,\theta,\zeta)$ and $Z(\rho,-\theta,-\zeta)=-Z(\rho,\theta,\zeta)$, halves these bases by admitting only the products of a poloidal and a toroidal factor that have the required parity under a simultaneous sign reversal of $\theta$ and $\zeta$. For the even function $R$ these are $\cos(m\theta)\cos(nN_{fp}\zeta)$ and $\sin(|m|\theta)\sin(nN_{fp}\zeta)$, so $R$ retains its $P$ pairs together with the $N+1$ cosine modes and its $Q$ pairs together with the $N$ sine modes, leaving $P(N+1)+QN$ coefficients. The odd functions $Z$ and $\lambda$ retain the opposite pairing and therefore $Q(N+1)+PN$ coefficients. Only the modes that survive at $\zeta=0$ enter the Poincaré constraint, which is the group paired with the cosines in each case,
\begin{equation}
    N^R_p = P \hspace{1cm} N^Z_p = N^\lambda_p = Q \hspace{1cm} N^R_b = 2MN+M+N+1 \hspace{1cm} N^Z_b = N^R_b - 1
\end{equation}
the last difference being the $(m,n)=(0,0)$ mode, which the sine basis does not contain.

The two conditions differ in the functions they act on. The prescribed LCFS constrains the geometry alone, so that $N_b = N^R_b + N^Z_b$, and leaves $\lambda$ free up to its gauge. The Poincaré condition constrains the same two functions and, optionally, $\lambda$ on the plane as well, as discussed in \S2. Summed over the constrained functions, the stellarator-symmetric totals are
\begin{equation}
    N_b = (2M+1)(2N+1) \hspace{1cm} N_p = P + Q \hspace{1cm} N_p^{\lambda \text{ fixed}} = P + 2Q
\end{equation}
For $L=M=12$ and $N=6$, the resolution used for the quasi-helical optimization of \S\ref{sec:stage1}, the LCFS thus contributes 325 rows, against 91 for the Poincaré condition, or 133 when the $\lambda$ cross-section is fixed as well. Per function, these are 49 rows against 163 for $R$, which has 595 coefficients in total, and 42 against 162 for $Z$, which has 588.

The Poincaré totals do not involve $N$, since the condition acts on a single plane, whereas the LCFS total grows linearly with the toroidal resolution. Equating them, with Equation \ref{PQ-counts} for $P$ and $Q$, gives
\begin{equation}
    N = \frac{L}{4} - \frac{M(M+2)}{4(2M+1)} \approx \frac{2L-M}{8} \hspace{1cm} N = \frac{2L(3M+1) - M(3M+2)}{8(2M+1)} \approx \frac{3(2L-M)}{16}
\end{equation}
without and with the $\lambda$ cross-section fixed, respectively, and dropping stellarator symmetry changes these estimates only marginally. Beyond such a toroidal resolution the Poincaré boundary condition is the less restrictive of the two, which for $L=M$ leaves only $N\lesssim M/8$ on the other side. Raising the radial resolution at fixed $M$ raises the threshold, and the heliotron of \S\ref{sec:eqsolve}, at $L=24$, $M=12$ and $N=3$, falls below it: the Poincaré condition constrains the geometry there with 241 rows against 175 for the LCFS, and the solve still converges to the lower force error reported in that section.

\section{Comparison of QH Optimization}\label{app:comparison}
\begin{figure}
    \centering
    \includegraphics[width=\linewidth]{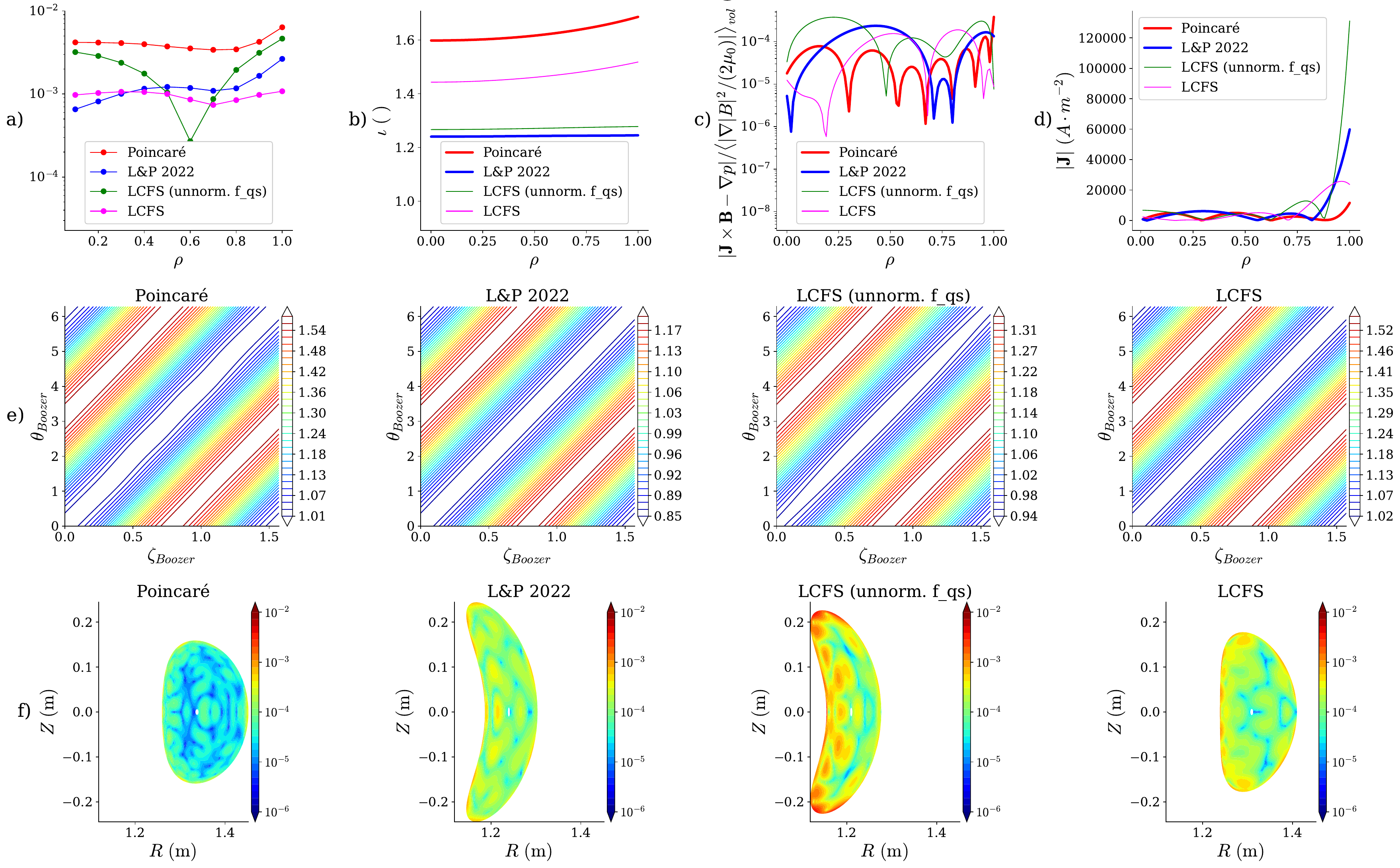}
    \caption{The four quasi-helically optimized configurations. As a function of the flux surface label: a) the quasisymmetry error $f_{qs}$, b) the rotational transform, c) the force error normalized by volume-averaged magnetic pressure, and d) the current density magnitude, the last two averaged over each surface. e) $|B|$ on the $\rho=1$ surface in Boozer coordinates, and f) the normalized force error on the $\zeta=0$ cross-section, which is the plane held fixed in the Poincaré case.}
    \label{compare-1}
\end{figure}

\begin{figure}
    \centering
    \includegraphics[width=0.8\linewidth]{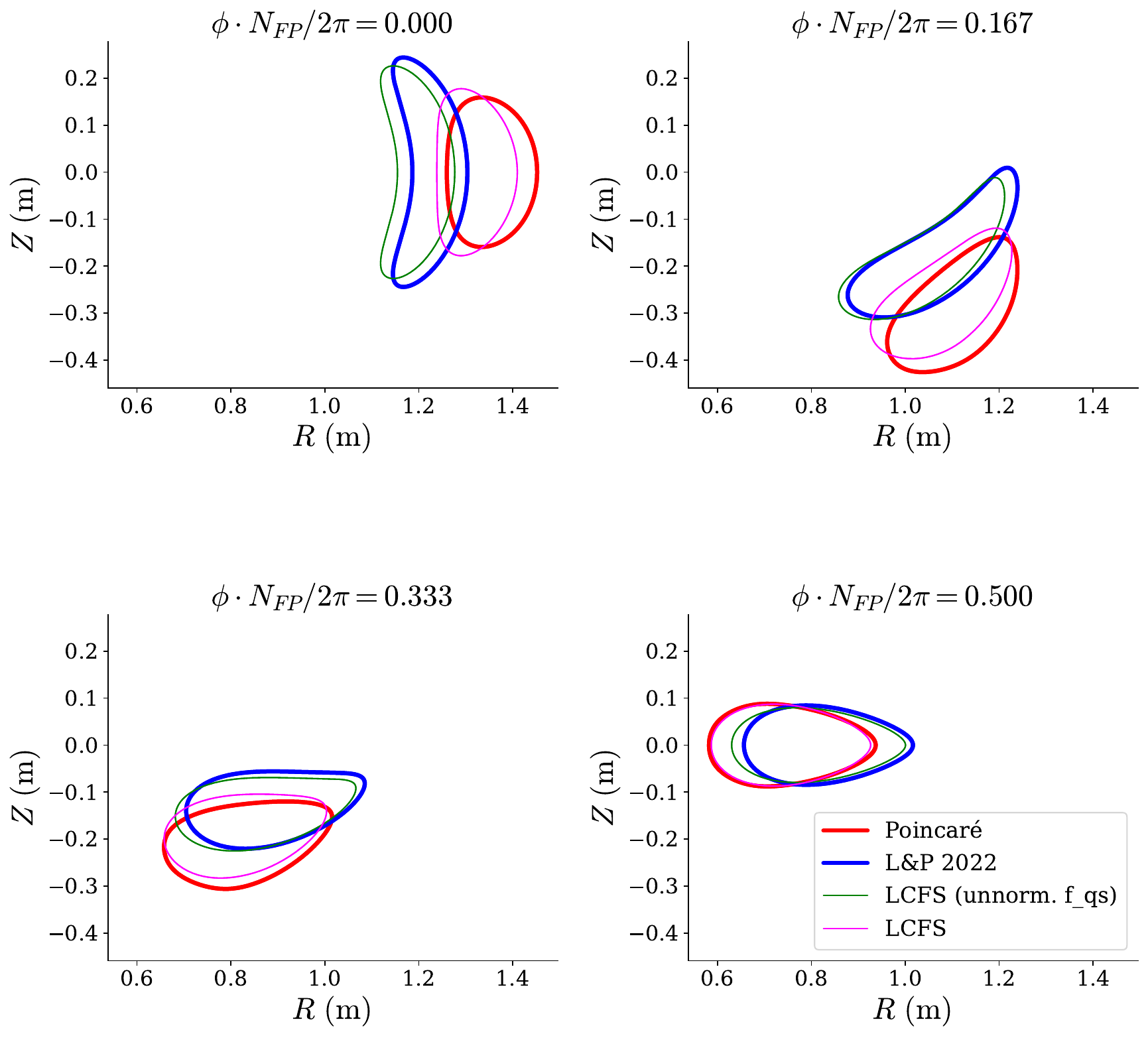}
    \caption{The last closed flux surface of the four configurations at four equispaced toroidal angles between $\zeta=0$ and $\zeta=\pi/N_{fp}$.}
    \label{compare-2}
\end{figure}

The optimization of \S\ref{sec:stage1} differs from a conventional LCFS optimization in multiple ways. The use of Poincaré variables and a fixed Poincaré section equilibrium solve is the main one, but we also add a $\langle |B| \rangle$ objective and normalize $f_{qs}$ by the volume-averaged field strength. To separate these effects, we compare four optimized equilibria. \textbf{Poincaré} is the result of \S\ref{sec:stage1}. \textbf{L\&P 2022} is the precise QH configuration of \cite{landremanMagneticFieldsPrecise2022}, imported into \texttt{DESC} and resolved, which carries no $\langle |B| \rangle$ target. \textbf{LCFS (unnorm. $f_{qs}$)} is optimized with the LCFS method and the $\langle |B| \rangle$ target, but without the normalization in $f_{qs}$. \textbf{LCFS} uses the same objectives as the Poincaré case and optimizes the LCFS boundary shape instead. The results of the four optimizations are shown in Figures \ref{compare-1} and \ref{compare-2}, and the corresponding global quantities are listed in Table \ref{tab:compare}. Except for \textbf{Poincaré}, all the equilibria have $LMN=8$ whereas \textbf{Poincaré} has $LM=12$ and $N=6$. We used a higher resolution for $LM$ for \textbf{Poincaré} since those are the optimized modes, in return, we reduced the toroidal resolution $N$. 

\begin{table}
  \centering
  \begin{tabular}{lccccc}
    \hline
    Configuration & $V$ ($\text{m}^3$) & $R_0/a$ & $\langle |B| \rangle$ (T)
      & $\langle |F| \rangle_{n}$ & $f_{qs}$ at $\rho=1$ \\
    \hline
    Poincaré                        & 0.307 & 8.0 & 1.2595 & 2.06E-5 & 6.35E-3 \\
    L\&P 2022                       & 0.306 & 8.0 & 0.9966 & 1.31E-4 & 2.64E-3 \\
    LCFS (unnorm. $f_{qs}$)         & 0.284 & 8.0 & 1.1117 & 4.11E-4 & 4.64E-3 \\
    LCFS                            & 0.295 & 8.0 & 1.2563 & 2.61E-4 & 1.08E-3 \\
    \hline
  \end{tabular}
  \caption{Global quantities of the four optimized configurations. $\langle |F| \rangle_{n}$ is the volume-averaged force error normalized by the volume-averaged magnetic pressure, as everywhere else in this paper. $R_0/a$ is the aspect ratio.}
  \label{tab:compare}
\end{table}


\textbf{Poincaré} and \textbf{LCFS} are very similar in shape, but the Poincaré-optimized case has a noticeably better force balance, by an order of magnitude as shown in Table \ref{tab:compare} against the otherwise identical LCFS run and by a similar amount against the other two. We note that the normalized force error improvement is independent of the equilibrium resolution, and this is validated by solving all the equilibria with the same resolution $LM=12$ and $N=8$ with the same tolerances using the fixed LCFS solve. We see that the relative improvement stays the same.

The quasisymmetry errors are comparable across the four, with the Poincaré case the worst of them. We attribute this to the choice of optimization hyperparameters rather than to the boundary condition. For the LCFS optimizations, we used weights, an initial trust radius and a resolution that previous studies have shown to work well, whereas the corresponding choices for the Poincaré run have not been refined to the same extent. We also kept the same initial guess throughout, which is known to affect the outcome substantially, and a different one may well close the gap.

The largest effect on the optimized boundary comes from the $\langle |B| \rangle$ target, and from how well it is met. With the total toroidal magnetic flux essentially the same in all four cases, targeting the volume-averaged field strength keeps the boundary close to the axial torsion of the initial guess. The spread in $\langle |B| \rangle$ is much larger than the differences in flux or minor radius, so it is the shaping that is responsible for it. The rotational transform, which none of the runs targeted, varies over a comparable range and follows the same grouping, from about 1.25 for \textbf{L\&P 2022} and \textbf{LCFS (unnorm. $f_{qs}$)} to about 1.6 for \textbf{Poincaré}. \textbf{L\&P 2022} carries no such target and settles at a lower field strength, while \textbf{LCFS (unnorm. $f_{qs}$)} has the target but, with the weights used, does not reach 1.26 T.

All the optimizations used vacuum profiles, yet the LCFS-optimized cases all carry a significant current close to the $\rho=1$ boundary, whereas the Poincaré case, which has the lowest force error of the four, stays nearly current-free there. In a perfect vacuum solution, the current has to be zero, and we take this residual current to be another symptom of the imperfect force balance rather than a separate effect. The four configurations do not order in the same way by edge current and by force error, so we do not read a quantitative relation into it. How much such a spurious current affects the quasisymmetry metric is not clear, and the differences here are small enough that all four remain usable. We nonetheless expect a solution with a lower force error to give the more trustworthy metric in general.

\bibliographystyle{jpp}
\bibliography{Poincare-Paper}

\end{document}